\documentclass[runningheads]{llncs}
\usepackage[T1]{fontenc}
\usepackage{graphicx}
\usepackage{booktabs}
\usepackage{pdflscape}
\usepackage{amsmath}
\usepackage{hyperref}

\usepackage{color}

\newcommand\blfootnote[1]{%
  \begingroup
  \renewcommand\thefootnote{}%
  \footnote{#1}%
  \addtocounter{footnote}{-1}%
  \endgroup
}
\makeatletter
\def\blfootnote{\gdef\@thefnmark{}\@footnotetext}
\makeatother

\begin{document}
\title{Setting the clock: Evaluating temporal window parameters for coordinated behavior detection}
\titlerunning{Evaluating temporal window parameters for coordinated behavior detection}

%%% GP: for now, removing orcids. To add again for the camera-ready submission?

\author{
Georgios Panayiotou
\inst{1}
%\orcidID{0009-0002-3907-3189}
\and
Lorenzo Mannocci
\inst{2}
%\orcidID{0000-0002-5556-3746}
\and
Maurizio Tesconi
\inst{3}
%\orcidID{0000-0001-8228-7807}
}
\authorrunning{Panayiotou et al.}
\institute{
%Infolab
InfoLab, Dept. of Information Technology, Uppsala University, Uppsala, Sweden \\
\email{georgios.panayiotou@it.uu.se}
\and
%UniPi
University of Pisa, Pisa, Italy \\
\email{lorenzo.mannocci@di.unipi.it}
\and
%iit-cnr
IIT-CNR, Pisa, Italy \\
\email{maurizio.tesconi@iit.cnr.it}
}
\maketitle              % typeset the header of the contribution

\begin{abstract}
Coordinated behavior is a central mechanism of online collective action. On social media platforms, it can support legitimate mobilization, but it can also be exploited in disinformation campaigns, astroturfing, and information operations. 
Detecting coordinated behavior on social media platforms typically relies on  coordination networks, where users are linked when they perform similar actions within shared temporal windows. 
While the temporal window is central to how coordination is operationalized, it is often treated as an implementation detail rather than as a substantive modeling decision. 
This paper presents a first analysis of how two key temporal parameters, window length and stride, affect the detection of coordinated communities within information operation campaigns. 
We find that window length determines which coordination patterns are detectable, while window stride has negligible effect on precision and recall.
Our analysis highlights selecting appropriate temporal window parameters as an open methodological challenge requiring careful treatment.

\end{abstract}
%
%
%

%% ack publishing
\blfootnote{Accepted as a full paper at the AIDEM Workshop, ECMLPKDD 2026.}

\section{Introduction}
\label{sec:introduction}

Coordination is a fundamental mechanism of online collective behavior. On social media platforms, users routinely coordinate to organize protests, sustain social movements, support public causes, amplify emergency information, or collectively discuss unfolding events~\cite{mannocci2024detection}. At the same time, similar coordination mechanisms can also be exploited for harmful purposes, including disinformation campaigns, astroturfing, online harassment, and information operations~\cite{cima2024coordinated}. Coordinated behavior is therefore not inherently malicious or inauthentic: rather, it describes a broad class of collective dynamics in which multiple actors perform mutually reinforcing actions toward an explicit or implicit goal~\cite{mannocci2024detection}. This broader view is important because the same observable patterns---for instance, many accounts sharing the same content or acting within a short time interval---may arise from both grassroots mobilization~\cite{nizzoli2021coordinated} and orchestrated manipulation~\cite{mannocci2022mulbot}.

The increasing relevance of coordinated behavior has motivated the development of several computational methods for its detection. A large part of this literature relies on network science approaches, which construct so-called \emph{coordination networks}, representing users as nodes and connecting them when they perform similar actions. Typical examples include co-retweeting the same post, sharing the same URL, using the same hashtag, or mentioning the same account~\cite{pacheco2021uncovering,weber2021amplifying,nizzoli2021coordinated}. Once such a network is built, community detection algorithms are commonly used to identify groups of users whose actions are more strongly aligned with one another than with the rest of the network~\cite{mannocci2024detection}.

A central modeling choice in this process is the definition of a valid \emph{co-action}. In many approaches, two users are considered coordinated only if they perform the same action on the same object within a given temporal window. Hence, time windows operationalize the notion of synchronization: they determine whether two similar actions are sufficiently close in time to be interpreted as evidence of coordination~\cite{weber2021temporal}. Prior work has adopted a wide range of temporal settings, using windows from seconds to minutes, hours, days, and weeks, and relying on different strategies such as adjacent, overlapping, or action-driven windows~\cite{mannocci2024detection}. However, these choices are often treated as implementation details rather than substantive modeling assumptions.

This is problematic because temporal windowing directly shapes the coordination network that is eventually analyzed. A very short window imposes a strict synchronization constraint and may reveal tightly orchestrated behavior, but it can also produce sparse graphs and miss slower forms of coordination. Conversely, a long window is more permissive and can capture looser or slower collective dynamics, but it may also connect users who independently interact with the same popular content over extended periods. Similarly, overlapping windows may reduce boundary effects, where temporally close actions fall into different adjacent windows, but they also increase computational cost and may not necessarily improve the quality of the detected communities. Despite these consequences, there is still limited empirical evidence on how window length and window stride affect the detection of coordinated communities.

%\paragraph{Contributions.} 
The present paper addresses this gap by treating temporal window selection as the main parameter of interest. Rather than assuming a fixed window size or overlap, we evaluate how different temporal configurations affect the construction of co-repost networks and the communities detected within them. We center our analysis around two research questions:
%% GP: limiting for space constraints
(i) How does temporal window length affect the detection of coordinated communities?
(ii) How does window stride, and therefore overlap between consecutive windows, affect detection performance?

To answer these questions, we construct temporal co-repost networks under multiple window lengths, ranging from seconds to one day, and compare non-overlapping windows with overlapping sliding windows. We then apply community detection to each temporal network and evaluate the resulting communities in terms of their structure and their agreement with labelled coordinated actors.

Our results show that temporal window length has a substantial effect on the detected communities. Short windows capture highly synchronized forms of coordination, while longer windows reveal slower and more diffuse patterns, but may also introduce noisier associations. In contrast, window overlap has a much weaker effect: across the tested configurations, overlapping windows do not consistently improve detection compared to adjacent non-overlapping windows. These findings suggest that temporal window length should be treated as a key modeling parameter in coordinated behavior detection, while overlap may often be unnecessary when the goal is to detect coordinated communities at scale.

%% GP: changed mentions of co-retweets to co-reposts, as the seckin et al paper does not explicitly name the source as twitter/x.

% can be cut if space is needed in the end
% \textbf{Paper organization.} The remainder of the paper is organized as follows. Section~\ref{sec:related_work} reviews related work on coordinated behavior detection, with a focus on network-science methods and temporal windowing. Section~\ref{sec:method} describes the construction of temporal co-retweet networks, the community-detection procedure, and the evaluation metrics. Section~\ref{sec:results} presents the empirical results across temporal window configurations. Section~\ref{sec:discussion} discusses the implications of our findings for coordinated-behavior detection. Finally, Section~\ref{sec:conclusion} concludes the paper and outlines future work.

\section{Related work}
\label{sec:related_work}

% Detecting good windows in temporal modularity-based clustering
A large body of work detects coordinated online behavior through network-based methods. These approaches typically construct a \emph{coordination network}, where nodes represent users and edges encode evidence that two users performed similar or identical actions. Examples of such co-actions include retweeting the same post, sharing the same URL, using the same hashtag, mentioning the same account, commenting on the same content, or publishing similar textual or visual material~\cite{nizzoli2021coordinated,pacheco2021uncovering,weber2021amplifying,tardelli2024multifaceted}. The underlying assumption is that coordination can be inferred from repeated, non-random similarities in user behavior. Differently from social or interaction networks, coordination networks can connect users even when they never directly interact with each other, because the link is induced by similarity of action rather than by an explicit social tie~\cite{weber2021amplifying,pacheco2021uncovering}.

Network science methods usually follow a common analytical pipeline~\cite{mannocci2024detection}. First, a set of relevant users is selected, often based on activity, content production, resharing behavior, or topic participation. Second, a coordination network is constructed by defining one or more co-actions and a similarity function between pairs of users. Edge weights may count the number of shared actions, such as common retweets or URLs, or use vector-based similarity measures such as cosine similarity over TF--IDF representations~\cite{cinelli2022coordinated,nizzoli2021coordinated,pacheco2021uncovering}. Third, the network is filtered in order to remove weak or noisy similarities. Finally, community detection algorithms such as Louvain~\cite{blondel2008fast} are applied to identify groups of users that exhibit stronger internal coordination than expected by chance~\cite{tardelli2024temporal}.
% removed ref to Leiden to save space

Within this framework, time is a crucial modeling dimension. Coordinated behavior is not only defined by users performing similar actions, but also by the temporal proximity of those actions. For this reason, many network-based methods introduce temporal constraints during network construction or filtering: two actions are considered a valid co-action only if they occur within the same temporal window~\cite{magelinski2022synchronized,pacheco2021uncovering,weber2021temporal}. Therefore, temporal window acts as a filter that operationalizes synchronicity. It determines whether behavioral similarity should be interpreted as evidence of coordination or as a weaker form of coincidental similarity.

Existing studies adopt different windowing strategies. Adjacent time windows partition the observation period into non-overlapping intervals, such as 15 minutes, 1 hour, 6 hours, 1 day, or 1 week~\cite{weber2020who,weber2021amplifying,vargas2020detection,cima2024coordinated}. Evenly distributed overlapping windows instead slide over time with a fixed step, so that consecutive windows partially overlap; prior work has used window sizes ranging from 1 second to several minutes, hours, or days~\cite{graham2020like,magelinski2020detecting,magelinski2022synchronized,pacheco2021uncovering,ng2022online,mannocci2026multimodal,tardelli2024multifaceted}. A third strategy is based on action-driven overlapping windows, where each window is positioned around the occurrence of a relevant action. This approach has been used with short windows, often between a few seconds and a few minutes, to capture tightly synchronized behavior~\cite{broniatowski2021towards,pacheco2020unveiling,giglietto2020takes}.

The choice of temporal window has direct methodological consequences~\cite{seiron2023modularitybased}. Adjacent windows are computationally simple, but they may miss pairs of actions that occur close in time but fall on opposite sides of a window boundary. Overlapping windows mitigate this boundary problem because temporally close actions have a greater chance of being included in at least one common window. However, they also increase the number of windows that must be processed, and therefore, the computational cost of the analysis~\cite{panayiotou2024current,seiron2023modularitybased}. Action-driven windows avoid fixed temporal partitions by centering the analysis around observed actions, but they define co-actions relative to the timing of specific events rather than to a uniform temporal grid~\cite{mannocci2024detection}.

Window length is equally important. Short windows impose a strict synchronization requirement and are therefore suited to detecting highly orchestrated behavior, where users act within seconds or minutes of each other~\cite{magelinski2022synchronized,pacheco2021uncovering}. Longer windows are more permissive and may capture slower or more weakly synchronized forms of coordination, including emergent or loosely organized collective behavior~\cite{nizzoli2021coordinated,tardelli2024temporal}. However, increasing the window length also increases the probability of linking users who independently interact with the same popular content, thereby introducing noisy edges and potentially reducing the specificity of the detected communities.

Despite the central role of time in defining co-actions, the choice of temporal window is often treated as a technical parameter. Prior studies have used highly heterogeneous window sizes, ranging from seconds to minutes, hours, days, and weeks, and have relied on different window types, including adjacent, evenly distributed overlapping, and action-driven overlapping windows~\cite{mannocci2024detection}, while others avoid discrete windows altogether, instead weighting co-actions continuously by temporal proximity using a decay function~\cite{iannucci2026detecting}. 
This heterogeneity suggests that there is no established consensus on the appropriate temporal scale for detecting coordinated behavior. Moreover, only a limited number of works explicitly investigate how temporal modeling choices affect the resulting coordination networks and detected communities~\cite{weber2021temporal,tardelli2024temporal,ng2022online}. 
Our work addresses this gap by evaluating how temporal window length and stride affect the detection of coordinated communities.
% ================================================================
\section{Method}
\label{sec:method}

We study how temporal windowing affects the detection of coordinated communities in co-repost networks. Following network-based approaches to coordinated behavior detection, we model coordination through \emph{co-actions}: two accounts are connected when they perform the same action on the same content within a given temporal context. In our setting, the action is a repost and the shared object is the reposted content. We represent the resulting structure as a temporal multilayer network, where each layer corresponds to a temporal window. We then detect communities by jointly considering consecutive temporal layers and evaluate how the detected communities vary under different temporal configurations.

\subsection{Temporal co-repost network}
\label{subsec:temporal_corepost_network}

%% GP: shortening less essential def:s equations into inlines

Let $U$ be the set of accounts, and $\mathcal{R}$ the set of repost actions observed in a campaign. A temporal configuration is defined by a window length $t_x$ and a stride $t_s$. The window length determines the duration of each temporal window, while the stride determines the temporal distance between the starting times of two consecutive windows. 
These parameters induce an ordered sequence of temporal windows $\mathcal{T} = (T_1,T_2,\ldots,T_m)$.
% \begin{equation}
% \mathcal{T} = (T_1,T_2,\ldots,T_m).
% \end{equation}
Each window $T_i$ has duration $t_x$. Consecutive windows start $t_s$ time units apart; therefore, $t_s=t_x$ yields adjacent non-overlapping windows, while $t_s<t_x$ yields overlapping windows with overlap $t_x-t_s$. 

We represent the resulting temporal coordination structure as a temporal multilayer co-repost network $G = (V,E,W,\mathcal{T})$,
% \begin{equation}
% G = (V,E,W,\mathcal{T}),
% \end{equation}
where $V \subseteq U$ is the set of accounts involved in at least one co-retweet, $\mathcal{T}$ is the set of temporal layers, $E$ is the set of intra-layer edges, and $W$ is the set of edge weights. 
Each temporal layer $T_i \in \mathcal{T}$ induces a weighted graph $G_i = (V_i,E_i,W_i)$,
% \begin{equation}
% G_i = (V_i,E_i,W_i),
% \end{equation}
where $V_i \subseteq V$ contains the accounts involved in co-reposts during $T_i$. An edge $(u,v) \in E_i$ is added when accounts $u$ and $v$ repost at least one common post within $T_i$. Its weight is the number of distinct posts co-reposted by the pair in that window:
\begin{equation*}
w_i(u,v) =
\left|
\{p : u \text{ and } v \text{ both reposted } p \text{ within } T_i\}
\right|.
\end{equation*}

Thus, each temporal layer captures co-repost coordination within a specific time interval. Edges encode behavioral similarity rather than social ties or direct interactions. Isolated nodes are discarded from each layer, since accounts with no co-action partners do not contribute to community detection.

\subsection{Community detection on temporal layers}
\label{subsec:community_detection}

To detect coordinated communities, we jointly consider consecutive temporal layers. Specifically, for each pair of consecutive layers $G_i$ and $G_{i+1}$, we construct a flattened graph:
\begin{equation*}
G_{i:i+1} = (V_{i:i+1},E_{i:i+1},W_{i:i+1}),
\end{equation*}
where the node set is the union of the nodes appearing in the two layers:
$
V_{i:i+1} = V_i \cup V_{i+1}.
$
The edge set is obtained by taking the union of the edges in $G_i$ and $G_{i+1}$. When the same edge appears in both layers, its weight in the flattened graph is the sum of the two layer-specific weights:
$
w_{i:i+1}(u,v) = w_i(u,v) + w_{i+1}(u,v),
$
where missing weights are treated as zero.
This approach allows the following detection step to account for co-reposting activity that spans nearby temporal windows. This is particularly relevant near window boundaries, where closely timed reposts may otherwise be separated into different temporal layers.

Community detection is then applied to each flattened graph $G_{i:i+1}$. This produces a partition
$
\mathcal{P}_{i:i+1} =
\{C_1^{i:i+1}, C_2^{i:i+1}, \ldots, C_{k_{i:i+1}}^{i:i+1}\},
$
where each $C_j^{i:i+1}$ is a detected coordinated community. 

\subsection{Evaluation metrics}
\label{subsec:evaluation_metrics}

We evaluate each temporal configuration by characterizing the communities detected across all flattened graphs. Since the datasets used for our evaluation include labelled accounts belonging to verified Information Operation (IO) campaigns, we can compare detected communities against a set of known coordinated actors. Note that these labels are used only for evaluation: they are not used during network construction or community detection.

Let $U_{IO} \subseteq U$ be the set of labelled IO actors in a campaign dataset. We evaluate the detected communities using both structural metrics and label-based metrics.

\paragraph{Structural properties.}
For each flattened graph $G_{i:i+1}$, we first compute the number of detected communities $n_c$. We then measure the size of each community in the corresponding partition $\mathcal{P}_{i:i+1}$ relative to the total number of unique accounts in the campaign:
\begin{equation}
s(C_j^{i:i+1}) = \frac{|C_j^{i:i+1}|}{|U|}.
\end{equation}
This normalization allows us to compare community sizes across campaigns with different numbers of accounts. For each temporal configuration, we aggregate these quantities across all flattened graphs, obtaining the mean number of detected communities $\bar{n}_{c}$ and the mean relative community size $\bar{s}$.

\paragraph{Agreement with labelled IO actors.}
For each detected community $C_j^{i:i+1}$, we compute IO precision as:
\begin{equation}
\mathrm{IO~precision}(C_j^{i:i+1}) =
\frac{|C_j^{i:i+1} \cap U_{IO}|}{|C_j^{i:i+1}|}.
\end{equation}
This measures the fraction of community members that are labelled IO actors. High IO precision indicates that the detected community is mostly composed of known coordinated actors.

We also compute IO recall at the flattened-graph level:
\begin{equation}
\mathrm{IO~recall}(G_{i:i+1}) =
\frac{
\left|
\bigcup_{C_j^{i:i+1} \in \mathcal{P}_{i:i+1}}
(C_j^{i:i+1} \cap U_{IO})
\right|
}{
|U_{IO}|
}.
\end{equation}
This measures the fraction of labelled IO actors that appear in at least one detected community in $G_{i:i+1}$. High IO recall indicates that a temporal configuration captures many known coordinated actors, although not necessarily within the same community.

Precision and recall capture complementary effects of temporal windowing. Short windows impose a stricter synchronization constraint and may produce high-precision communities when labelled actors act in close temporal proximity, but they may miss slower or less synchronized actors. Longer windows are more permissive and may increase recall, but they can also connect labelled IO actors with other users who repost the same content over longer periods, reducing precision.

\paragraph{Community cohesion.}
Finally, we measure the internal cohesion of each detected community as the mean weight of its intra-community co-repost edges. For a community $C_j^{i:i+1}$, cohesion is defined as:
\begin{equation}
\omega(C_j^{i:i+1}) =
\frac{1}{|E_C|}
\sum_{\substack{u,v \in C_j^{i:i+1} \\ (u,v) \in E_{i:i+1}}}
w_{i:i+1}(u,v),
\end{equation}
where $E_C = \{(u,v) \in E_{i:i+1} : u,v \in C_j^{i:i+1}\}$ is the set of intra-community edges.
This metric captures the frequency of co-actions among accounts in the same community. Higher values indicate that community members repeatedly co-repost the same content within the considered temporal context. For each temporal configuration, we aggregate cohesion values across communities and flattened graphs, obtaining the mean intra-community cohesion $\bar{\omega}$.

% ================================================================
\section{Experimental setting}
\label{sec:experimental_setting}

%% GP: converting to paragraphs to save space

% \subsection{Datasets}
% \label{subsec:datasets}

\paragraph{Datasets}
We consider 14 datasets representing a variety of verified IO campaigns on online social network platforms~\cite{seckin2025labeled}.
These datasets contain ground-truth information about accounts that are part of the respective IO campaigns. This makes them suitable for evaluating coordinated community detection, because detected communities can be compared against labelled campaign actors. Importantly, the IO labels are used as a ground-truth proxy for known coordinated actors, not as a restriction of the method to information operations only. In other words, the detection method remains general, while the labelled IO datasets provide a controlled benchmark to assess how different temporal settings affect the recovery of known coordinated actors.
%% GP: adding note for reviewer :-)
For brevity, we refer to~\cite{seckin2025labeled} for more details on data collection and IO-actor composition of these campaigns.

The presence of labelled actors also allows us to characterize the composition of the detected communities. Communities with high IO precision are mostly composed of labelled campaign actors, while communities with lower IO precision may include non-labelled users who reposted the same content within the same temporal context. This distinction is useful for understanding whether a temporal configuration isolates known campaign actors or merges them with broader user activity.

% \subsection{Temporal configurations}
% \label{subsec:experimental_temporal_configurations}

\paragraph{Temporal configurations}
We instantiate the temporal windowing method with two families of configurations. First, we use adjacent non-overlapping windows, where the stride is equal to the window length, i.e., $t_s=t_x$. We test the following window lengths:
$
t_x \in \{10\text{sec}, 1\text{min}, 15\text{min}, 1\text{h}, 6\text{h}, 12\text{h}, 1\text{day}\}.
$
Second, we use overlapping sliding windows with the same window lengths. For each $t_x$, we test two stride values:
$
t_s \in \{0.2t_x, 0.5t_x\}.
$
These correspond to $80\%$ and $50\%$ overlap between consecutive windows, respectively. This design allows us to evaluate separately the role of window length and stride. Window length controls the temporal scale at which reposts are considered synchronized, while stride controls the degree of overlap between consecutive temporal layers.

% \subsection{Community detection implementation}
% \label{subsec:community_detection_implementation}

\paragraph{Community detection}
For each dataset and temporal configuration, we construct the corresponding co-repost network and flatten each pair of consecutive temporal layers as described in Section~\ref{subsec:community_detection}. We then apply the Louvain community detection algorithm~\cite{blondel2008fast} to each flattened graph, as it is an oft-used community detection method in prior work. For every temporal configuration, we compute the structural metrics, IO precision, IO recall, and community cohesion defined in Section~\ref{subsec:evaluation_metrics}. Metrics are first computed at the flattened-graph level and then aggregated across the campaign. This produces one set of summary statistics for each dataset and each temporal configuration, allowing us to compare how window length and overlap affect the detection of coordinated communities.
\section{Results}
\label{sec:results}

We evaluate the effect of temporal windowing along three dimensions. First, we describe how window length affects the structural properties of the detected communities. Second, we measure how well the detected communities recover labelled IO actors. Third, we analyze whether overlapping windows improve detection compared to adjacent non-overlapping windows. 

\subsection{Effect of window length on community structure}
\label{subsec:results_structure}
For each temporal configuration, we characterize the detected communities using three structural metrics: the average number of detected communities $\bar{n}_c$, the average relative community size $\bar{s}$, and the average cohesion $\bar{\omega}$. The first metric measures how many communities are detected on average in each temporal window; the second measures community size as a percentage of campaign accounts; and the third measures the average intra-community co-repost weight.

\begin{table}[tb!]
  \centering
  \caption{Community statistics per campaign and window size for non-overlapping windows. Values are averaged across all temporal windows. $\bar{n}_{c}$: mean number of detected communities; $\bar{s}$: mean community size as a percentage of campaign accounts; $\bar{\omega}$: mean intra-community cohesion. Dashes (---) indicate that the temporal length of the campaign is smaller than the window length.}
  \label{tab:structural_stats}
  \scalebox{0.57}{
  \setlength{\tabcolsep}{4pt}
  \begin{tabular}{l rrr rrr rrr rrr rrr rrr rrr}
    \toprule
    & \multicolumn{3}{c}{\textit{10 sec}} 
    & \multicolumn{3}{c}{\textit{1 min}} 
    & \multicolumn{3}{c}{\textit{15 min}} 
    & \multicolumn{3}{c}{\textit{1 h}} 
    & \multicolumn{3}{c}{\textit{6 h}} 
    & \multicolumn{3}{c}{\textit{12 h}} 
    & \multicolumn{3}{c}{\textit{1 day}} \\
    \cmidrule(lr){2-4} 
    \cmidrule(lr){5-7} 
    \cmidrule(lr){8-10} 
    \cmidrule(lr){11-13} 
    \cmidrule(lr){14-16} 
    \cmidrule(lr){17-19} 
    \cmidrule(lr){20-22}
    \textbf{campaign} 
    & $\bar{n}_{c}$ & $\bar{s}$ & $\bar{\omega}$ 
    & $\bar{n}_{c}$ & $\bar{s}$ & $\bar{\omega}$ 
    & $\bar{n}_{c}$ & $\bar{s}$ & $\bar{\omega}$ 
    & $\bar{n}_{c}$ & $\bar{s}$ & $\bar{\omega}$ 
    & $\bar{n}_{c}$ & $\bar{s}$ & $\bar{\omega}$ 
    & $\bar{n}_{c}$ & $\bar{s}$ & $\bar{\omega}$ 
    & $\bar{n}_{c}$ & $\bar{s}$ & $\bar{\omega}$ \\
    \midrule
    \texttt{armenia} & 1.4 & 0.29 & 0.85 & 1.9 & 0.27 & 1.11 & 4.6 & 0.27 & 1.18 & 8.3 & 0.27 & 1.22 & 27.9 & 0.29 & 1.31 & 59.8 & 0.31 & 1.53 & 77.0 & 0.34 & 1.82 \\
    \texttt{bangladesh} & 1.0 & 1.95 & 0.98 & 1.0 & 1.51 & 1.11 & 1.9 & 0.59 & 1.49 & 3.4 & 0.47 & 1.51 & 10.6 & 0.37 & 2.09 & 16.8 & 0.37 & 2.19 & 27.0 & 0.38 & 1.97 \\
    \texttt{catalonia} & 1.8 & 0.14 & 0.84 & 2.5 & 0.17 & 1.23 & 6.8 & 0.25 & 2.14 & 18.1 & 0.26 & 2.36 & 68.6 & 0.30 & 2.30 & 106.3 & 0.34 & 1.74 & 101.0 & 0.32 & 1.41 \\
    \texttt{china-1} & 2.9 & 0.01 & 1.78 & 3.8 & 0.02 & 2.44 & 12.9 & 0.02 & 3.02 & 38.2 & 0.02 & 3.32 & 188.9 & 0.03 & 4.26 & 352.6 & 0.03 & 4.52 & 576.0 & 0.03 & 4.87 \\
    \texttt{china-2} & 1.4 & 0.03 & 1.61 & 1.8 & 0.02 & 2.13 & 8.3 & 0.02 & 3.58 & 27.6 & 0.02 & 3.93 & 145.5 & 0.02 & 4.53 & 270.7 & 0.02 & 4.60 & 475.3 & 0.02 & 5.13 \\
    \texttt{ecuador} & 2.3 & 0.06 & 2.39 & 3.1 & 0.05 & 3.43 & 10.3 & 0.03 & 6.28 & 30.1 & 0.03 & 4.99 & 145.3 & 0.03 & 4.51 & 287.8 & 0.03 & 2.34 & 538.0 & 0.04 & 2.32 \\
    \texttt{egypt\_uae} & 1.5 & 1.54 & 2.11 & 2.3 & 1.75 & 4.58 & 3.0 & 1.76 & 12.19 & 3.8 & 1.50 & 21.15 & 5.6 & 1.50 & 28.12 & 7.0 & 1.68 & 30.97 & 14.0 & 1.97 & 1.46 \\
    \texttt{ghana\_nigeria} & 1.3 & 0.36 & 1.04 & 1.9 & 0.30 & 1.72 & 4.3 & 0.32 & 3.13 & 9.5 & 0.32 & 3.26 & 29.3 & 0.31 & 2.17 & 42.8 & 0.31 & 1.81 & 83.0 & 0.38 & 2.33 \\
    \texttt{qatar} & 1.6 & 0.02 & 1.39 & 2.8 & 0.02 & 2.39 & 13.2 & 0.04 & 4.03 & 38.7 & 0.04 & 4.08 & 167.1 & 0.05 & 3.62 & 286.2 & 0.06 & 3.32 & 525.7 & 0.06 & 4.03 \\
    \texttt{spain} & 3.6 & 0.34 & 1.86 & 7.0 & 0.56 & 3.70 & 40.6 & 0.63 & 5.47 & 93.0 & 0.70 & 3.51 & --- & --- & --- & --- & --- & --- & --- & --- & --- \\
    \texttt{thailand} & 2.9 & 0.16 & 1.35 & 4.2 & 0.30 & 2.05 & 7.9 & 0.56 & 2.23 & 15.4 & 0.57 & 2.32 & 42.3 & 0.93 & 2.25 & 70.0 & 1.25 & 1.48 & --- & --- & --- \\
    \texttt{uae} & 7.5 & 0.11 & 1.72 & 13.7 & 0.09 & 1.83 & 11.6 & 0.09 & 3.03 & 22.1 & 0.10 & 4.37 & 74.8 & 0.08 & 3.64 & 147.0 & 0.06 & 2.77 & 282.0 & 0.08 & 2.93 \\
    \texttt{venezuela-1} & 2.5 & 0.06 & 1.06 & 4.1 & 0.09 & 1.42 & 14.2 & 0.18 & 1.77 & 34.0 & 0.21 & 1.53 & 104.5 & 0.32 & 1.40 & 180.0 & 0.35 & 2.08 & 196.0 & 0.38 & 1.98 \\
    \texttt{venezuela-2} & 1.3 & 0.15 & 3.80 & 1.9 & 0.14 & 7.36 & 7.9 & 0.16 & 25.42 & 18.5 & 0.16 & 22.19 & 66.1 & 0.17 & 1.40 & 95.5 & 0.18 & 0.98 & 191.5 & 0.21 & 1.46 \\
    \bottomrule
  \end{tabular}
  }
\end{table}

Table~\ref{tab:structural_stats} reports these metrics for non-overlapping windows across all campaigns and window sizes. The clearest pattern is that the average number of detected communities increases with window length in almost every campaign. This is expected: longer windows contain more repost actions, creating more opportunities for accounts to be connected through shared reposts.

This increase in community count is not accompanied by a comparable increase in relative community size. Across most campaigns, $\bar{s}$ remains relatively stable as the window grows. This suggests that larger windows do not simply merge users into a few large communities. Instead, they tend to reveal a larger number of communities with similar proportional size. In other words, increasing the temporal window mostly increases the amount of detectable co-action structure, rather than collapsing the network into larger aggregate groups.

Cohesion also tends to increase with window length, although less uniformly than community count. This follows from the definition of $\omega(\cdot)$: longer windows give pairs of accounts more opportunities to co-repost the same content multiple times, increasing intra-community edge weights. Campaigns such as \texttt{egypt\_uae} and \texttt{venezuela-2} show particularly large cohesion values at intermediate window sizes, indicating repeated co-reposting within detected communities. Other campaigns show more moderate variation, suggesting that additional co-action opportunities do not always translate into substantially stronger intra-community ties.
Overall, Table~\ref{tab:structural_stats} shows that temporal window length has a clear structural effect on the detected communities. Larger windows make more coordinated structure visible, but they do not necessarily produce larger communities. This supports the idea that window length controls the temporal scale at which coordination becomes observable.

\subsection{Effect of window length on IO precision}
\label{subsec:results_window_length_precision}

\begin{figure}[t!]
\centering
\includegraphics[width=\linewidth]{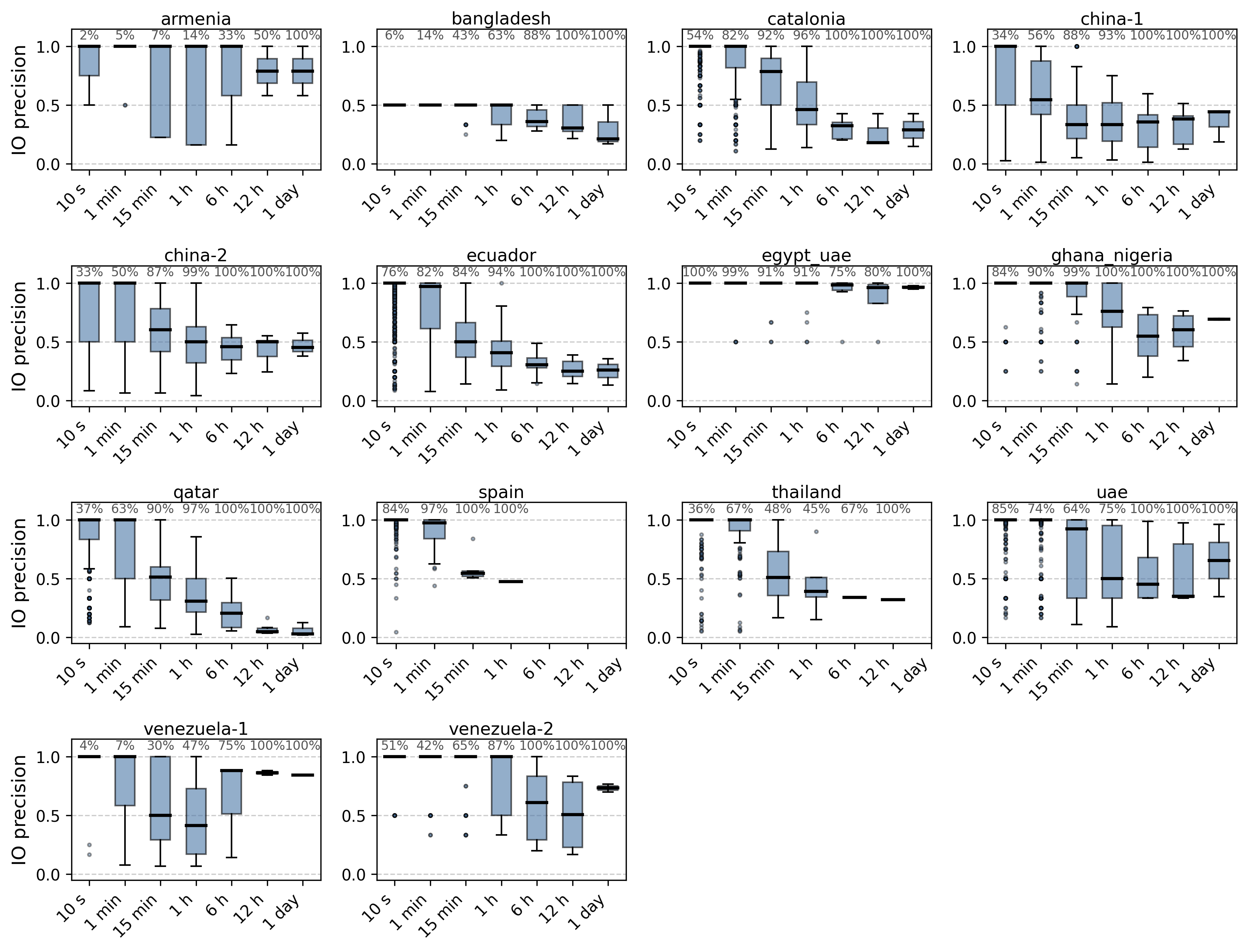}
\caption{Conditional IO precision by window size $t_x$. Precision is computed over windows in which at least one IO-majority community is detected; the activity rate (top of each panel) denotes the fraction of windows of each $t_x$ setting where this condition holds.}
\label{fig:size_precision}
\end{figure}
Figure~\ref{fig:size_precision} reports the distribution of IO precision computed on windows where at least one IO-majority community is detected. The figure also shows the corresponding activity rate, defined as the fraction of windows in which such a community appears. For the smallest window sizes, IO-majority communities are often detected with very high precision. Precision then generally decreases as the window length grows, suggesting that larger windows progressively include users outside the labelled IO set. However, high precision at fine temporal resolutions does not always come with low activity rate. Some campaigns, such as \texttt{egypt\_uae} and \texttt{ghana\_nigeria}, show both high precision and activity rates above $80\%$ even for small windows, indicating frequent and tightly synchronized coordination. In contrast, campaigns such as \texttt{armenia} and \texttt{venezuela-1} show near-zero activity rates at small window sizes, meaning that highly precise IO-majority communities rarely appear at fine temporal resolutions. Most campaigns fall between these two cases, reflecting the different scales and nature of coordination between the various campaigns.

Overall, small windows are effective at capturing tightly synchronized coordinated communities when they occur, but they may miss campaigns with long inactive periods or slower coordination rhythms. Larger windows are more likely to detect coordinated communities across the campaign, but they increasingly merge labelled IO actors with organic users who repost the same content over longer time spans.

\subsection{Effect of window stride on detection performance}
\begin{figure}[t!]
\centering
\includegraphics[width=\linewidth]{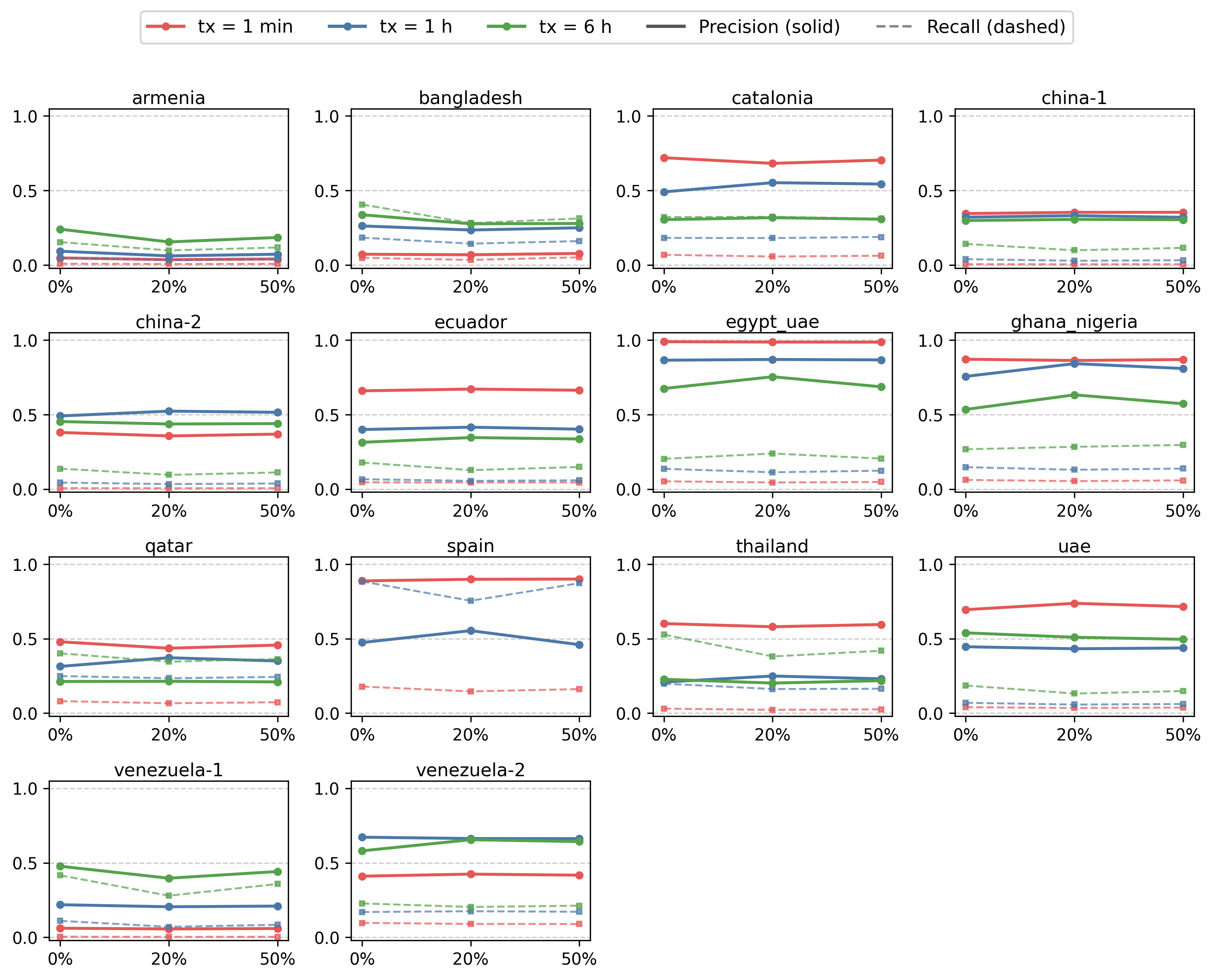}
\caption{Effect of window stride $t_s$ on IO precision (solid lines) and recall (dashed lines) for various window sizes, shown as a percentage of the window size $t_x$.}
\label{fig:stride_precision}
\end{figure}

Figure~\ref{fig:stride_precision} shows how IO precision and IO recall vary as the window stride increases from $0\%$ to $50\%$ of the window length, for three representative window sizes. Across campaigns, both metrics remain largely stable: changing the stride produces only small variations in either precision or recall. This indicates that increasing the overlap between consecutive windows does not substantially improve the detection of coordinated communities in our setting.

These results suggest that window stride plays a much smaller role than window length. While overlapping windows can, in principle, reduce boundary effects between adjacent windows, we do not observe a consistent gain in detection performance. 
Therefore, non-overlapping windows appear sufficient for this task, with the additional advantage of requiring fewer temporal graphs to construct and analyze.

\subsection{IO precision and cohesion capture different signals}
\label{subsec:results_precision_cohesion}

\begin{figure}[t!]
\centering
\includegraphics[width=\linewidth]{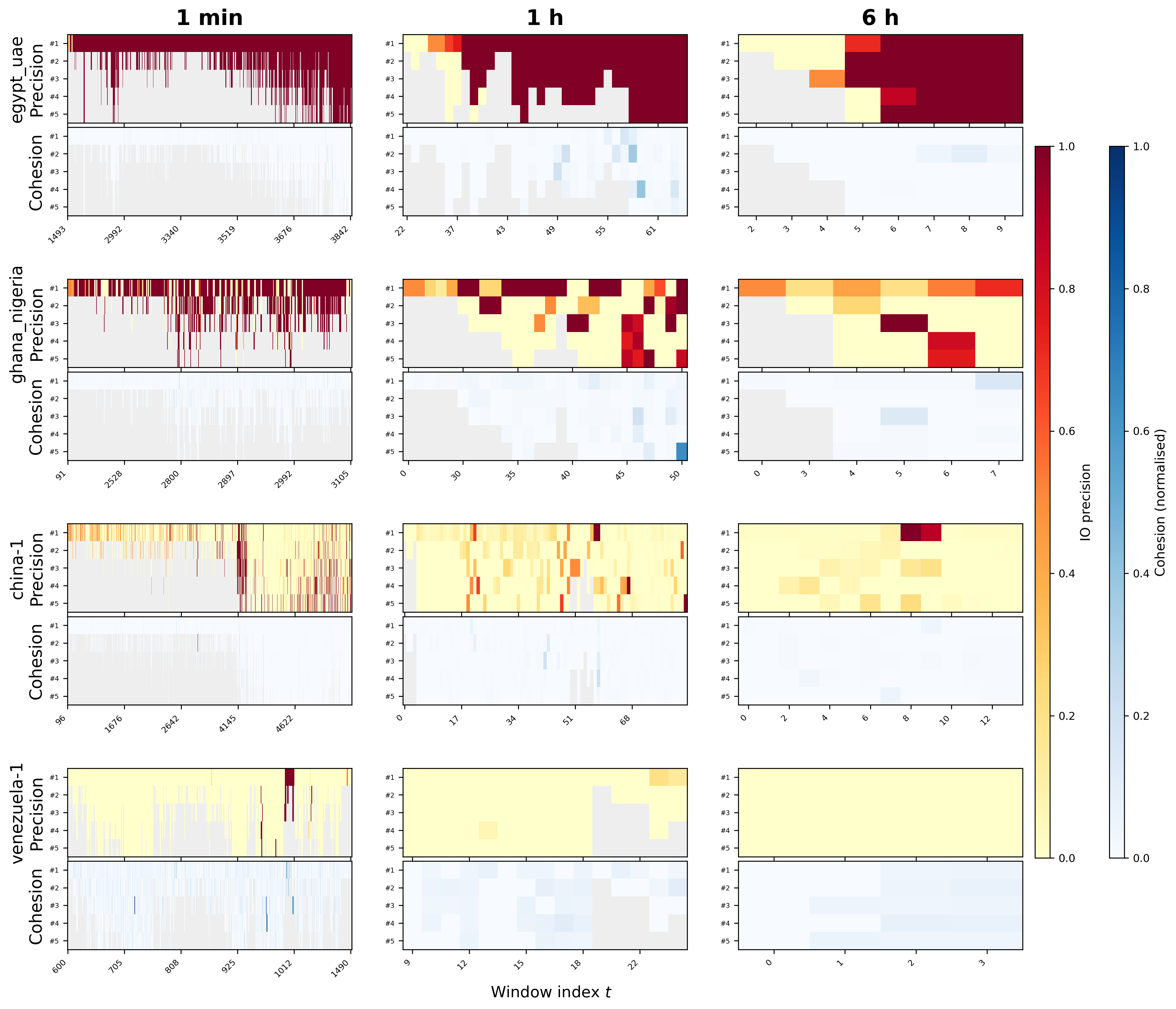}
\caption{Per-window IO precision (top) and intra-community cohesion (bottom) for the top-5 largest communities across all non-overlapping windows, for three window sizes (1 min, 1 h, 6 h).}
\label{fig:top5_precision}
\end{figure}

Finally, Figure~\ref{fig:top5_precision} compares IO precision and normalized cohesion for the five largest communities in each window, focusing on four representative campaigns with different numbers of active users and different durations.

The main result is that IO precision and community cohesion are largely decoupled. IO precision captures the composition of a community, i.e., whether its members are mostly labelled IO actors. Cohesion, instead, captures the intensity of repeated co-reposting within the community. Therefore, a community can be highly precise without necessarily being highly cohesive.

The \texttt{egypt\_uae} campaign provides a clear example. At 1 minute windows, the largest community reaches near-perfect IO precision across most of the campaign. However, cohesion remains low, suggesting that labelled IO actors frequently appear together in the same communities but rarely co-repost the same content more than once within each short window. When the window length increases to 1 hour, cohesion becomes more visible in specific periods, as repeated co-reposting has more time to accumulate.

The same separation between precision and cohesion is visible, albeit with different temporal patterns, in the other campaigns. For \texttt{china-1} and \texttt{venezuela-1} especially, IO precision appears in more localized spikes, while cohesion also varies over time. Overall, these results show that community composition and behavioral intensity should be interpreted as complementary dimensions of coordinated behavior.
\section{Discussion}
\label{sec:discussion}

%% GP: converting subsections to paragraphs to save space
% \subsection{RQ1: Effect of temporal window length}
% \label{subsec:discussion_rq1}

\paragraph{Effect of temporal window length}
Our results show that temporal window length substantially affects the detection of coordinated communities. This effect, however, is not uniform across campaigns. In some cases, such as campaigns characterized by frequent and tightly synchronized co-reposting, high-precision IO communities are visible even with very short windows. In other cases, fine-grained windows detect few or no IO-majority communities, and coordination becomes visible only when longer windows are used.

This suggests that window length determines which temporal scale of coordination is observable. Short windows emphasize strict synchronization and are therefore well suited to detecting rapid, high-frequency coordination. Longer windows are more permissive and can reveal slower or more intermittent forms of coordination, but they also increase the likelihood of connecting labelled IO actors with non-labelled users who repost the same content over longer periods. The choice of window length is therefore not a neutral implementation detail: it changes the type of coordinated behavior that the network representation can expose.

The relationship between IO precision and cohesion further clarifies this point. Our results show that these two quantities capture different properties of detected communities: a community can have high IO precision while showing low cohesion, meaning that labelled actors appear together in the same community without repeatedly co-reposting the same content within the window. This pattern is consistent with a form of parallel coordination, where accounts act on the same targets in a synchronized way but without dense repeated co-actions. Conversely, a community with high cohesion and lower IO precision may indicate a highly co-acting group that also includes organic or non-labelled participants. Both cases are relevant, but they correspond to different coordination patterns and should not be conflated.

% \subsection{RQ2: Effect of window stride}
% \label{subsec:discussion_rq2}

\paragraph{Effect of window stride}
Our second research question concerns the role of window stride, and therefore the amount of overlap between consecutive temporal windows. Across the tested configurations, changing the stride produces only small variations in IO precision and IO recall. In particular, overlapping windows do not consistently improve the recovery of labelled IO actors, nor do they produce communities with higher IO precision.

This result suggests that, for the co-repost networks and community detection procedure considered here, window length is more important than window overlap. Overlapping windows can in principle reduce boundary effects, since actions occurring close in time are less likely to be separated into different windows. However, this potential benefit does not translate into a measurable improvement in our experiments. Adjacent non-overlapping windows therefore appear sufficient for detecting coordinated communities in this setting.
%% GP: adding note here about overlap and flattening
We note, however, that our pipeline flattens consecutive layer pairs prior to community detection, a step which itself can partially mitigate boundary effects between adjacent windows. 
% It is therefore possible that this flattening step partly explains the limited effect of stride observed in this study.
To generalize these results, future studies should consider more extensive benchmarking, including additional community detection methods that can consider multiple temporal network layers. 
% as well as comparisons of different strategies of accounting for actions close to the borders of the temporal windows.

Still, this finding has some practical implications, considering that overlapping windows substantially increase the number of temporal graphs that must be constructed and analyzed. Since community detection can be computationally expensive, especially when extended to larger datasets or richer network representations, avoiding unnecessary overlap reduces computational cost without sacrificing detection performance. This is particularly important for scalable or near-real-time coordinated behavior detection pipelines.

\section{Conclusion}
\label{sec:conclusion}

This paper investigates how temporal windowing affects the detection of coordinated communities in co-repost networks. Using labelled information operation datasets as ground truth, we evaluate how different window lengths and strides influence the structure of detected communities and their agreement with known coordinated actors. Our results show that window length is a central modeling choice: short windows can isolate highly synchronized and high-precision communities, while longer windows reveal slower forms of coordination, but increasingly mix labelled IO actors with other users. In contrast, window stride has a limited effect on IO precision and recall, suggesting that overlapping windows provide little benefit over adjacent non-overlapping windows for this task. 
%% GP: shortening also paragraph below
Overall, our findings highlight how temporal windowing should not be treated as a minor parameter, as the chosen window length determines which temporal scale of coordination becomes visible.
%
% Our findings highlight that temporal windowing should not be treated as a minor implementation detail. Instead, the chosen window length determines which temporal scale of coordination becomes visible, and therefore shapes the conclusions drawn from coordination network analysis. At the same time, our results suggest that non-overlapping windows may be sufficient in many settings, reducing the computational cost of temporal coordinated community detection.
%

%% GP: add paragraph on limitations -- things that should be done in extended work, that reviewers also point out.

% Limitations & future work: (1) multimodal
As a first step,
% towards general guidance on temporal windowing in coordinated community detection
our evaluation is focused on a single mode of co-actions (reposts). Coordinated behavior, however, can involve multiple co-actions and modalities, including shared URLs, hashtags, mentions, images, and videos. Recent work has shown the importance of multimodal and multilayer representations for capturing richer coordination signals~\cite{mannocci2026multimodal,wohlert2026detecting}; an important next step is to test whether the effects noted in this work generalize to settings beyond co-repost networks, where each layer may follow a different temporal rhythm.

% (2) beyond louvain and flattening
Moreover, our pipeline is constrained to a specific community detection method, while it also flattens consecutive temporal layer pairs before community detection, which may partly absorb part of the benefit that overlapping windows would otherwise provide. 
Extensions of this work should also compare alternative community detection algorithms, since different methods make different assumptions about community structure and may respond differently to sparse or dense temporal graphs, and further test detection on non-flattened layers to isolate the effect of stride from that of flattening.
Finally, temporal windowing should be studied together with other network construction choices, such as superspreader filtering, minimum co-action thresholds, edge-weight filtering, and user selection, to better understand how parameter choices behind the methodology affect the detection of coordinated communities.

% % Future work
% With the previous in mind, future work should extend this analysis beyond co-repost networks. Coordinated behavior can involve multiple co-actions and modalities, including shared URLs, hashtags, mentions, images, and videos. Recent work has shown the importance of multimodal and multilayer representations for capturing richer coordination signals~\cite{mannocci2026multimodal,wohlert2026detecting}; an important next step is to test whether the effects of window length and stride generalize to these settings, where each layer may follow a different temporal rhythm. 
% Further work should also compare alternative community detection algorithms, since different methods make different assumptions about community structure and may respond differently to sparse or dense temporal graphs. Finally, temporal windowing should be studied together with other network construction choices, such as superspreader filtering, minimum co-action thresholds, edge-weight filtering, and user selection, to better understand how parameter choices behind the methodology affect the detection of coordinated communities.

\begin{credits}
\subsubsection{\ackname}
This work was partly funded by the European Union's Horizon 2020 research and innovation programme under grant agreement No. 871042 (SoBigData++); by eSSENCE, an e-Science collaboration funded as a strategic research area of Sweden; by the European Union under the scheme HORIZON-INFRA-2025-01-DEV-02 -- Early phase implementation of ESFRI Projects that entered the ESFRI Roadmap in 2021, Grant Agreement No. 101292277, ``SoBigData IP: SoBigData Implementation Phase''; and by the HORIZON Europe projects TANGO -- Grant Agreement No. 101120763.
The authors are also grateful to Matteo Magnani for insightful feedback on early versions of the manuscript. 

\subsubsection{\discintname} The authors declare no competing interests.

%% GP: Collapsing to one to save some space
% \subsubsection{Data Availability.} The datasets analyzed are available at \url{https://doi.org/10.5281/zenodo.14141549}.
% \subsubsection{Code Availability.} The scripts supporting the analysis are available at (link to repository removed for peer review).

\subsubsection{Data Availability.} The datasets analyzed are publicly available on Zenodo (\url{https://doi.org/10.5281/zenodo.14141549}); see also Ref. \cite{seckin2025labeled}. The analysis scripts are available on Github (\url{https://github.com/giorgospanay/sd-tcb-validation}).

\end{credits}

\bibliographystyle{splncs04}
\bibliography{mybib} %references
% Note to self: here to avoid extra information, export from zotero to better bibtex

% add otehrs manually here if needed

\end{document}